\documentclass[nofootinbib,twocolumn,preprintnumbers,superscriptaddress]{revtex4-2}

\usepackage{amsmath,amssymb}
\usepackage{graphicx}
\usepackage{hyperref}
\usepackage{physics}
\usepackage{xcolor}
\usepackage{xspace,slashed}
\usepackage{qcircuit}
\usepackage{multirow}
\usepackage[utf8]{inputenc}
\usepackage{comment}
\usepackage{amsthm}
\usepackage{lipsum}
\usepackage{siunitx}
\usepackage{soul} 
\usepackage{bookmark}
\usepackage{braket}
\usepackage[capitalize]{cleveref}
\usepackage{ifthen}
\usepackage{textgreek}
\usepackage[T1]{fontenc}
\usepackage[inline]{enumitem}

\begin{document}

%

\title{Superconducting penetration depth of
    Aluminum thin films }

\newcommand{\BSC}{Barcelona Supercomputing Center, Barcelona 08034, Spain}
\newcommand{\ICCUB}{Departament de F\'isica Qu\`antica i Astrof\'isica and Institut de
    Ci\`encies del Cosmos, Universitat de Barcelona, Barcelona 08028, Spain}
\newcommand{\IFAE}{Institut de F\'\i sica d'Altes Energies, The Barcelona Institute of Science and Technology, Bellaterra 08193, Spain}
\newcommand{\Qilimanjaro}{Qilimanjaro Quantum Tech SL, Barcelona, Spain}
\newcommand{\CNM}{Institute of Microelectronics of Barcelona (IMB-CNM), Spanish National Research Council (CSIC), Cerdanyola 08193, Spain}
\newcommand{\UAB}{Departament de F\'isica, Universitat Aut\`onoma de Barcelona, 08193 Bellaterra, Spain}

\author{David~L\'opez-N\'u\~nez}
\affiliation{\IFAE}
\affiliation{\ICCUB}
\affiliation{\BSC}

\author{Alba~Torras-Coloma}
\affiliation{\IFAE}
\affiliation{\UAB}

\author{Queralt~Portell-Montserrat}
\affiliation{\IFAE}
\affiliation{\CNM}
\affiliation{\UAB}

\author{Elia~Bertoldo}
\affiliation{\IFAE}

\author{Luca~Cozzolino}
\affiliation{\IFAE}

\author{Gemma~Rius}
\affiliation{\CNM}

\author{M.~Martínez}
\affiliation{\IFAE}
\affiliation{\Qilimanjaro}

\author{P.~Forn-Díaz}
\email{pforndiaz@ifae.es}
\affiliation{\IFAE}
\affiliation{\Qilimanjaro}

%
\begin{abstract}
    We present a study of the superconducting penetration depth $\lambda$ in aluminum thin films of varying thickness. The range of thicknesses chosen spans from the thin-film regime to the regime approaching bulk behaviour. The penetration depths observed range from $\lambda = 163\pm1~\rm{nm}$ for the thinnest 28~nm samples down to $\lambda = 54\pm1~\rm{nm}$ for the 207~nm-thick ones, allowing us to provide an estimate of the thickness at which aluminum becomes a type-I superconductor. In order to accurately determine $\lambda$, we performed complementary measurements using the frequency of superconducting $LC$ resonators obtained through novel and efficient methods of fitting and simulation, as well as the normal-state resistance of meandered structures. Both methods yield comparable results, providing a well-characterized set of values of $\lambda$ in aluminum in the relevant range for applications in fields such as quantum computing and microwave radiation detector technologies.

\end{abstract}

\maketitle
%
\section{Introduction}

Bulk aluminum has been an extensively studied material in the literature~\cite{cochran}, often considered as the archetype of conventional type-I superconductivity. In contrast, the properties of thin-film aluminum, conventionally utilized in, e.g., quantum computing technologies~\cite{brooks, kim, blais, adrian} and kinetic inductance detectors~\cite{Coiffard2020}, are not as widely apparent. Specifically, thin-film aluminum exhibits type-II superconductivity~\cite{nsanzineza}, resulting in a complete different behavior than its bulk version.

In type-II superconductors, the short range of Cooper pairs results in the emergence of normal-metal paths, leading to the appearance of vortices within the material~\cite{tinkham}. Furthermore, type-II superconductors may exhibit a significantly large kinetic inductance, $L_k$, comparable or larger than their geometric inductance, $L_g$~\cite{watanabe, forn-diaz}. Accurately determining $L_k$ is important as it needs to be taken into account, e.g., in the design of quantum circuits~\cite{forn-diaz, Weber2017}, requiring a thorough material characterization, particularly concerning the superconducting penetration depth, $\lambda$. This parameter is intrinsically material-dependent and represents the distance a parallel magnetic field penetrates into a superconductor. External factors such as the presence of impurities, the thickness of the film, or the direction of the magnetic field lead to an effective value of $\lambda$ \cite{degennes}. In particular, the film thickness strongly influences microscopic properties such as the electrical resistivity and the electron mean-free path~\cite{mayadas1, mayadas2, mayadas3}, which then modify the value of $\lambda$. To the best of our knowledge, the relationship between thickness and penetration depth in aluminum in the range of thicknesses targeted in our study has not been extensively documented in the existing literature \cite{reale}.

In this work, we determine the penetration depth of thin-film aluminum samples across film thicknesses in the range 28~nm--207~nm, with the goal of providing a well-characterized set of values of $\lambda$ for thin-film aluminum usable in the design of, e.g., resonating structures or circuits to build qubits~\cite{adrian, blais, forn-diaz}. This determination is accomplished through two complementary methods: measurement of the resonance frequency of a number of lumped-element superconducting resonators, and by evaluation of the resistance of thin-film test structures. These methodologies align with previous investigations conducted on other superconducting materials, such as Nb~\cite{gubin}, NbN~\cite{yoshida1995} and granular aluminum~\cite{rotzinger}. In addition to $\lambda$, both methods combined allow us to characterize $L_k$ for all thicknesses studied. Therefore, the results obtained in this work and the methods developed are an important resource in the design of thin-film aluminum-based circuits. In order to properly evaluate $\lambda$, we have developed a new fitting method for resonator devices with accurate error estimation, as well as a simulation technique using an EM solver to efficiently account for finite thickness effects in the estimation of the inductance. Finally, our investigation yields an estimate of the range of film thicknesses at which the transition occurs between the type-II superconductivity exhibited by thin-film aluminum and the type-I superconductivity characteristic of vortex-free, bulk aluminum. This transition may be relevant for superconducting qubits~\cite{bylander, wang2014} and resonators~\cite{nsanzineza} since the presence of vortices alters the quality of devices made of thin-film aluminum.

The outline of this work is as follows: Section~\ref{sec:theory} introduces the theoretical aspects of superconductivity needed for the data analysis. In section~\ref{sec:mod}, the methods used in this work to extract the penetration depth $\lambda$ are described. Section~\ref{sec:sim} presents the simulation methods along with the device characteristics. Section~\ref{sec:exp} focuses on the experimental implementation. The experimental results are discussed in Sec.~\ref{sec:res}. Finally, Sec.~\ref{sec:conclussions} presents the conclusions of this work.

%
\section{Theoretical Background}
\label{sec:theory}

\begin{figure}[!hbt]
    \includegraphics[,width=1\linewidth]{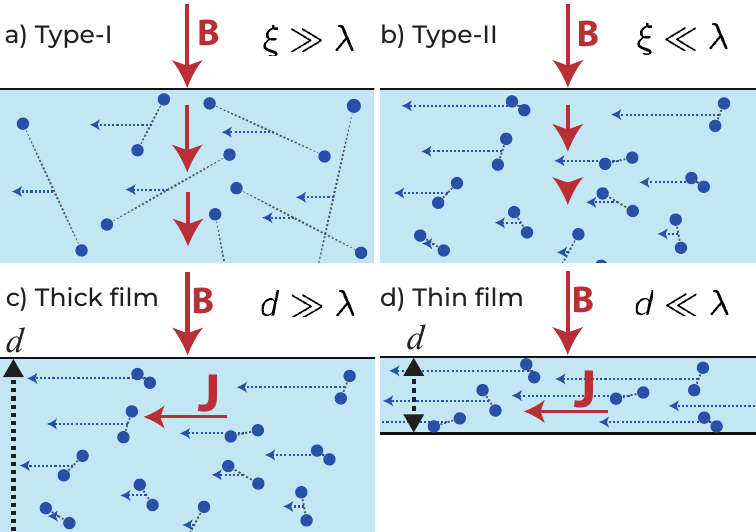}
    \caption{Qualitative representation of supercurrent behaviour under an applied magnetic field depending on superconductor type and thickness. \textbf{a)} Type-I superconductors, where response to an external perpendicular magnetic field is non-local. Each Cooper pair reacts to the magnetic field at its center of mass, since the average Cooper pair size $\xi$ is much larger than the typical distance at which the magnetic field decays, $\lambda$. This effectively reduces the field response, allowing the magnetic field to penetrate further into the material, thus increasing the penetration depth. \textbf{b)} In type-II superconductors, $\xi$ is smaller than $\lambda$, and the response is local, since both electrons of the Cooper pair are very close. \textbf{c)} Schematic representation of a local thick superconductor of thickness $d$ with $d\gg\lambda$, where the London equations are valid. The current response to the magnetic field follows $\Vec{j}_S = - \frac{\mu_0}{\lambda^2 } \Vec{A}_S$. \textbf{d)} In very thin films where $d\lesssim\lambda$, the Cooper pair number density is lower, and each Cooper pair has to increase its velocity to provide the same supercurrent $\Vec{j}_S$ as in thicker films, since the London equations are still valid. The kinetic inductance increases dramatically in this situation, as it is proportional to the squared velocity, $L_k\propto v^2$.}
    \label{fig:the-bac}
\end{figure}

In general, the superconductivity type exhibited by a given material is an expression of the interplay between the superconducting penetration depth $\lambda$ and its coherence length $\xi$, which corresponds to the average distance between the electrons forming a Cooper pair~\cite{pippard}.

The thickness of a thin film is manifested in the values of its internal microscopic parameters, as variations of both \(\lambda\) and \(\xi\) and the rest of parameters that depend on them. In particular, the kinetic inductance $L_k$ has a strong dependence on $\lambda$ through~\cite{rotzinger, mattis-bardeen, kautz, turneaure}
\begin{equation}
    \label{eq:kin-sq-pen}
    L_k = L_{k,s} N= \mu_0 \lambda N \, ,
\end{equation}
where $L_{k,s}$ is the surface kinetic inductance, or kinetic inductance per square. $N$ is the number of squares of a given wire. For wires with constant cross section, we have \(N=g/w\), where \(g\) and \(w\) are the wire length and width, respectively. Otherwise, $N$ has to be calculated along the wire.

In bulk aluminum, the response of the superconductor to magnetic fields is non-local [an intuitive explanation can be seen in Fig.~\ref{fig:the-bac}a)], causing $\lambda$ to be larger than the one predicted by the London equations $\lambda_L=15.7~\rm{nm}$. Since the Al bulk coherence length is $\xi_0=1.6~\unit{\um}\gg\lambda_L$, bulk aluminum is in the non-local limit. In the non-local limit, $\lambda$ is modified by~\cite{degennes}
\begin{equation}
    \lambda = 0.65(\lambda_L^2\xi_0)^{1/3},
\end{equation}
which, in the case of bulk aluminum, it is $\lambda_{\rm{bulk}}=50~\rm{nm}$~\cite{degennes, faber}. Thus, bulk aluminum satisfies type-I superconductivity criteria, $\xi_0>\lambda_{\rm{bulk}}$.

However, in polycrystalline aluminum, decreasing the film thickness $d$ eventually decreases the grain size as well\cite{meservey-2}, thereby reducing the electron mean free path, $l$. In this regime, an effective coherence length $\xi<\xi_0$ at $T=0~\unit{\K}$ can be defined~\cite{stjames, degennes}
\begin{equation}
    \label{eq:coh-len}
    \frac{1}{\xi} = \frac{1}{\xi_0} + \frac{1}{l} \,,
\end{equation}
which, in the dirty superconductor limit $\xi_0\gg l$, also modifies $\lambda$ as
\begin{equation}
    \label{eq:pen-dep-dirty}
    \lambda = a\lambda_L\sqrt{\frac{\xi_0}{l}} \, .
\end{equation}
$a$ is of order unity which depends on the surface scattering type, being $4/3\approx1.33$ for diffusive scattering and $\sqrt{4/3}\approx1.15$ for specular reflection~\cite{tinkham}. Equation~(\ref{eq:pen-dep-dirty}) is a reasonable approximation, particularly when $l\sim d$. Thus, by lowering $d$, one may arrive at the regime where $\xi<\lambda$, causing a change of superconductivity to a type-II superconductor, where the response to external fields is local [see \cref{fig:the-bac}b)]. The transition between superconductivity regimes is determined by the Ginzburg-Landau parameter $\kappa$~\cite{tinkham}
\begin{equation}
    \label{eq:kappa}
    \kappa \equiv \lambda / \xi.
\end{equation}
Above $\kappa=1/\sqrt{2}$, the superconductor is type-II since the normal-superconductor interface energy becomes negative, thus allowing for the generation of normal paths (vortices)~\cite{abrikosov}. For thick enough films, $\kappa$ decreases below $1/\sqrt{2}$ and the film exhibits vortex-free behavior, thus entering the type-I superconductivity regime. By carefully engineering the thicknesses of different thin films, this boundary between superconductivity types occurring at a critical thickness $d_c$ may be attained. In actual thin films, local fluctuations in $\kappa$ are expected given the typically disordered nature of the film, thus leading to a spread of values of $d_c$ over which the transition takes place.

For very thin films where $d\lesssim \lambda$ the response of the film to external perpendicular magnetic fields follows the London equations, causing an opposing current $\Vec{j}_S$ to an incoming field $\Vec{A}_S$. Since $d$ is so small, the magnetic field fully penetrates the film. Nevertheless, the expression $\Vec{j}_S = - \frac{\mu_0}{\lambda^2 } \Vec{A}_S$ remains valid~\cite{pearl}. In this scenario, the supercurrent response to $\Vec{A}_S$ is the same regardless of the film thickness [Fig.~\ref{fig:the-bac}c)]. However, as thinner films contain a lower amount of charge carriers, Cooper pairs are compelled to attain higher velocities, resulting in a substantial increase in $L_k$ (see Fig.~\ref{fig:the-bac}d), since $L_k \propto v^2$. The correction to $L_{k,s}$ due to this effect is given by~\cite{kautz}
\begin{equation}
    \label{eq:kin-ind-thin}
    L_{k,s} = \mu_0 \lambda \coth{\frac{d}{\lambda}} \equiv \mu_0 \lambda_{\text{thin}} \, ,
\end{equation}
where $\lambda_{\text{thin}}$ is an effective penetration depth. For $d\gg\lambda$, $\lambda_{\text{thin}}\to\lambda$, while for $d \ll\lambda$, $\lambda_{\text{thin}}\to\frac{\lambda^2}{d}$, significantly enhancing $L_{k,s}$. Resonator measurements, as will be detailed in the next section, give direct access to $L_{k}$, which can be used to derive $\lambda_{\rm{thin}}$ through Eq.~(\ref{eq:kin-ind-thin}), and then converted to $\lambda$.

Finally, $\lambda$ can be related to the normal metal properties of the thin film as it ultimately depends on the charge carrier properties, and can be extracted from similar principles as the Drude model. Specifically, in the dirty limit with $\xi_0,\lambda\gg l$, $\lambda$ is related to the normal state resistivity \(\rho_n\) of the metal at \(T = 0~\unit{K}\), through \cite{Hake1967}
\begin{equation}
    \label{eq:4probe-pen-dep}
    \lambda \simeq 105~\textrm{nm}\sqrt{\frac{\rho_n(\textrm{\textmu\textOmega}~\cdot\textrm{cm})}{T_c(\unit{K})}} \, ,
\end{equation}
where \(T_c\) is the superconducting critical temperature. Therefore, the value of $\lambda$ obtained through Eq.~(\ref{eq:kin-ind-thin}) may be validated by resistance measurements on thin films of the same nominal thickness $d$.

Equations~(\ref{eq:kin-ind-thin}, \ref{eq:4probe-pen-dep}) allow a complete characterization of $\lambda$ and $L_k$ of thin superconducting films, and are the basis of the experiment in this work.

%
\section{Methodology to extract $\lambda$}
\label{sec:mod}

\begin{figure}[!hbt]
    \centering
    \includegraphics{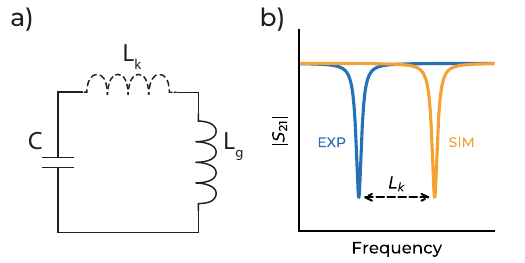}
    \caption{\textbf{a)} $LC$ Resonator circuit schematic. In the simulations conducted, no kinetic inductance is considered, while in the experiment the kinetic inductance $L_k$ contribution adds to the geometric inductance, $L_{\rm exp} = L_k + L_g$. \textbf{b)} Simulated and experimental resonances are assumed to only deviate due to $L_k$.}
    \label{fig:model}
\end{figure}

Following from the previous section, two independent approaches are here proposed and implemented to estimate $\lambda$ in aluminum thin films. The first method involves resonator measurements combined with EM simulation, where the effect of $\lambda$ is reflected in the measured resonance frequency. The second approach utilizes four-probe measurements on resistive aluminum meanders.

The resonator measurements give access to $\lambda$ through \(\lambda_{\mathrm{thin}}\), while the resistance measurements give direct access to \(\lambda\). Fabrication inaccuracies affect both methods differently. Discrepancies between nominal and real dimensions are much more difficult to track in resonators than in four-probe designs. \(\lambda_{\mathrm{thin}}\) is also much more sensitive to deviations in the thickness $d$. Resistance measurements are, thus, more robust. However, we include resonators in this work as they represent a good benchmark of a real-case scenario. 
Therefore, the resistance measurements in this study are in fact used to validate the resonator measurements.

\subsection{Resonance Method}
\label{ssec:mod-res}

Lumped element $LC$ resonators were chosen for this study in order to achieve a more controlled definition of their inductance $L$ and capacitance $C$, compared to distributed resonators. $LC$ resonators are characterized by their resonance frequency $\omega = (LC)^{-1/2}$, where $C$ represents the capacitance and $L = L_g + L_k$ is the sum of geometric and kinetic inductances. As explained in \cref{sec:theory}, superconductors may exhibit a significantly high $L_k$, especially for thin films with $d\ll\lambda$ \cite{forn-diaz}.

In order to extract $L_k$, we first perform an accurate simulation of each resonator studied considering a perfect conductor, thus obtaining $L_g$ and $C$ that lead to a simulated resonance frequency $f_{\mathrm{sim}}$ [\cref{fig:model}a) with \(L_k=0\)]. Resonator measurements provide $f_{\mathrm{meas}}$ as they contain a total inductance $L = L_g + L_k$ [\cref{fig:model}a)], so $f_{\mathrm{sim}}>f_{\mathrm{meas}}$ [\cref{fig:model}b)]. $L_k$ may be then obtained with
\begin{equation}
    \label{eq:lk-exp}
    L_k = L_g \left( \frac{f_{\mathrm{sim}}^2}{f_{\mathrm{meas}}^2} - 1\right) \,.
\end{equation}

The $L_k$ obtained in this way determines \(\lambda_{\mathrm{thin}}\) using \cref{eq:kin-ind-thin}. This method requires a precise simulation of the perfect conductor resonance $f_{\rm{sim}}$ (see Sec.~\ref{sec:sim}).

\subsection{Resistivity Method}
\label{ssec:mod-dcm}

Resistance measurements directly lead to $\lambda$ through Eq.~(\ref{eq:4probe-pen-dep}). Since $\rho_n$ is the normal state resistivity at $0~\unit{K}$, it can only be measured by destroying the superconducting state with an external magnetic field. However, the aluminum resistivity is nearly constant at low temperatures, so we instead approximate $\rho_n$ with the value at $4~\unit{K}$~\cite{cochran}.

It is important to note that $\rho_n$ is connected to $\lambda$ and not to \(\lambda_{\mathrm{thin}}\) as in the resonator method. \(\rho_n\) and \(\lambda\) are connected through the intrinsic properties of the material, such as the charge carrier mass, the Cooper pair number density, and the electron mean free path. Accordingly, the thin-film effects described in Eq.~(\ref{eq:kin-ind-thin}) do not play a role in the determination of $\lambda$ through $\rho_n$.

%
\section{Resonator simulations and device characteristics}
\label{sec:sim}

\begin{figure}[!hbt]
    \centering
    \includegraphics{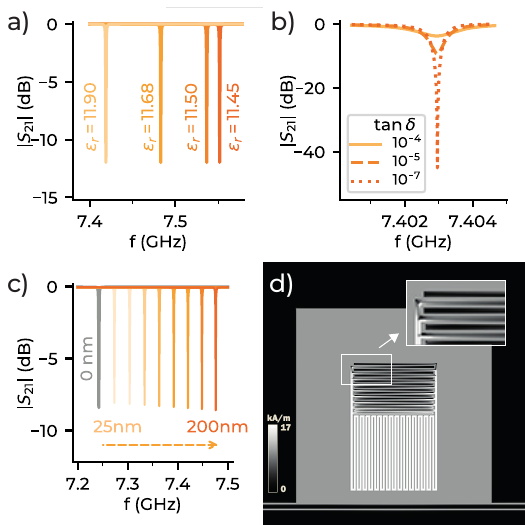}
    \caption{ \textbf{a)} $LC$ resonator simulated response with varying Si relative permittivity, $\epsilon_r$. Range of values is typically found in the literature. The actual value used in this work is $\epsilon_r = 11.45$~\cite{krupka}. \textbf{b)} Effect of Si loss tangent, $\tan \delta$, on the quality factor of $LC$ resonators. \textbf{c)} Sonnet simulations for an S resonator transmission. Grey curve represents the default response with no thickness. Yellow and increasing saturation color curves correspond to thicknesses ranging from $25~\unit{nm}$ to $200~\unit{nm}$. \textbf{d)} Simulated current distribution of a $LC$ resonator. The inset shows the current in the last capacitor fingers, where it is minimal.}
    \label{fig:sim}
\end{figure}

In this section, we summarize the procedure we followed to perform simulations on the resonator circuits, followed by the device characteristics.

First, for each film thickness $d$, we employ FastHenry\footnote{We use the version found in http://www.wrcad.com/ftp/pub/fasthenry-3.0wr-071720.tar.gz which takes into account the London equations and thus the penetration depth $\lambda$ in the inductance calculations.} to obtain an estimate of the kinetic inductance fraction \(\alpha\equiv L_k/L\). We work in the range $\alpha = 0.02-0.5$, as lower $\alpha$ can incur in big inaccuracies in $\lambda$, while higher $\alpha$ leads to frequencies lower than our experimental system bandwidth ($4-8~\unit{GHz}$). These simulations allow us to determine the inductor wire width $w$ in each resonator for a given $d$. 

Next, we employ Sonnet\footnote{https://www.sonnetsoftware.com/} to run an entire resonator simulation to obtain the system transmission (see \cref{app:sim} for details on the simulations). In these simulations we use the value of the electrical permittivity $\epsilon_r=11.45$, validated in ref.~\cite{krupka} which was also the low-temperature value confirmed by the wafer manufacturer. As seen in \cref{fig:sim}a), the value of the resonance changes noticeably for values of $\epsilon_r$ found throughout the literature~\cite{bruno, goppl, dimas, weber2011}, leading to a difference in the obtained $\lambda$ of up to $100\%$. Adding realistic values of substrate loss $\tan(\delta)$ to the simulations did not change the resonance frequency appreciably [see \cref{fig:sim}b)].

To discern the simulated values of $L_g$ and $C$ we follow the procedure described in ref.~\cite{doyle}.
A first simulation of the system yields the resonance frequency $f_{\rm sim}$. The values of $L_g$ and $C$ are estimated with an additional simulation wherein a known sheet inductance ($L_s$) is manually incorporated in a single square at the center of the inductor, leading to a modified resonance frequency $f_{\rm sim}'=1/(2\pi)\cdot[(L_g+L_s)C]^{-1/2}$. This procedure is equivalent to what is shown in \cref{fig:model}b), but with a manually incorporated \(L_k\).

However, in actual films a larger thickness $d$ decreases $L_g$ combined with an increase in $C$. This thickness-dependence effect cannot be directly captured by Sonnet since it is a 2.5D EM simulation software. To obtain the effect of $d$ on $L_g$ and $C$ in the simulation without resorting to slower, full 3D methods, the thickness is artificially simulated by introducing an additional aluminum layer above the existing structure and connecting both through vias along all edges. For the largest $d$ considered ($200~\mathrm{nm}$) the resulting correction due to thickness on the resonance frequency amounts to approximately $2\%$ [\cref{fig:sim}c)]. We have actually validated this method against a full 3D model. More details can be found in Appendix~\ref{app:sim}.

\begin{figure}[!hbt]
    \centering
    \includegraphics{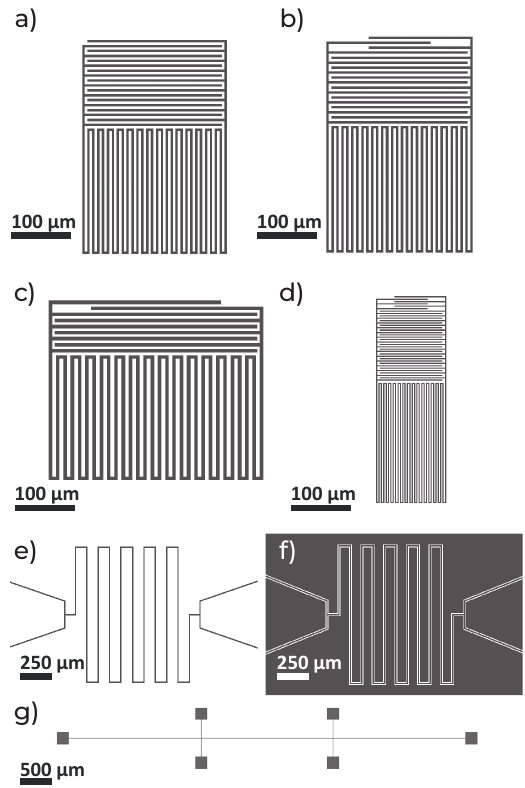}
    \caption{$LC$ resonator device design variations: (\textbf{a}) medium; (\textbf{b}) medium with adjusted finger capacitor; (\textbf{c}) large, (\textbf{d}) and small. Design of a 4-probe measurement meander without (\textbf{e}) and with (\textbf{f}) surrounding ground plane. (\textbf{g}) Layout of a Hall-bar measurement.}
    \label{fig:dev-des}
\end{figure}

Three types of resonators were designed with varying line widths and line gaps to introduce different contributions of $L_k$. The three designs are named small resonators (SR), medium resonators (MR), and large resonators (LR), each with a meander and capacitor widths of $2$, $4$, and $6~\unit{\um}$, respectively [See \cref{fig:dev-des}a)--d)]. LR are suited for higher $L_k$ contribution while SR are suited for larger thickness.

The lumped description with isolated \(L\) and \(C\) is justified with the current simulations [\cref{fig:sim}d)], where most of the current is present in the meander. Using current simulations as the one shown in Fig.~\ref{fig:sim}d), we estimate that 98\% of the energy is located inside the inductor meander. The extra \(2\%\) is taken into account to obtain an effective length of the meander for computing the number of squares $N$ in \cref{eq:kin-sq-pen}. Moreover, the ground plane is designed sufficiently far from the resonator so we can neglect its contribution to the inductance.

In order to have better statistics in the value of $\lambda$ for a given thickness, resonators with the same $L$ and different $C$ are designed. This is obtained by changing the outermost fingers of the capacitor [Figs.~\ref{fig:dev-des}a)-d)], where there is nearly no contribution to $L$ [see inset of \cref{fig:sim}d)]. In \cref{fig:dev-des}b) an example of a MR with the last fingers modified is shown. This finger modification reduces $C$, thus increasing the resonance frequency. The resonators are separated enough in frequency space to be experimentally distinguishable ($>25~\unit{MHz}$) in the range $7-8~\unit{GHz}$.

For the four-probe measurements, a long meander structure was designed [see \cref{fig:dev-des}e) and f)]. Two different meanders were considered, with and without ground plane around the meander. The size of the meander was chosen to ensure that the resistance at $4~\unit{\K}$ falls within a measurable range of $100~\unit{\ohm} - 1~\unit{\kohm}$. A Hall-bar structure has also been used [see \cref{fig:dev-des}g)] as a complementary measurement of the resistance.

The thicknesses $d$ chosen for the study range from \(25~\unit{\nm} \) to \(200~\unit{\nm} \). This range includes typical values found for superconducting qubit devices and radiation detectors (\(50~\unit{\nm} \)--\(100~\unit{\nm} \)). Including this range is indispensable for calibrating \(\lambda\) for real-scenario devices. $d<50~\rm{nm}\sim\lambda_{\rm{bulk, Al}}$ is chosen to observe thin-film effects, while $d>100~\rm{nm}$ is chosen to explore the regime approaching bulk aluminum behavior. The actual thicknesses were accurately determined using Atomic Force Microscopy.

\begin{figure}[!hbt]
    \centering
    \includegraphics{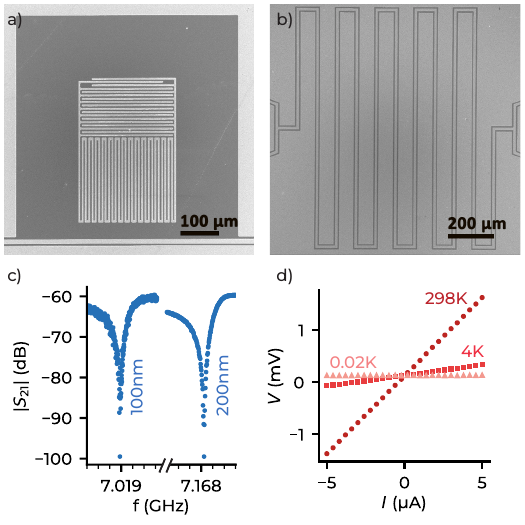}
    \caption{\textbf{a)} Scanning electron migrograph (SEM) of a $LC$ resonator. \textbf{b)} SEM of a meander structure measured by 4-probe method. \textbf{c)} $S_{21}$ data and fit of measured resonators with $100$~nm and $200$~nm thickness. Each minor tick corresponds to $1~\unit{MHz}$ \textbf{d)} Resistance measurements at room temperature (RT), $4~\unit{K}$ and $10~\unit{mK}$.}
    \label{fig:fab-exp}
\end{figure}

The devices were fabricated using photolitography on high resistivity silicon wafers with the Aluminum deposited by electron-beam evaporation. The chips were later diced and devices were wirebonded to a ceramic PCB, or to a chip carrier in the case of four-probe devices. Images from actual devices can be seen in Figs.~\ref{fig:fab-exp}a)-b). More details regarding thin film fabrication can be found in \cref{app:fab}, including images of three samples of increasing thickness, also displaying a clear increase of apparent grain size.

Two different evaporators, one Plassys at IFAE and another from Univex at CNM, were used in this study in order to account for possible fabrication dependencies on $\lambda$. While the IFAE evaporator is only used for aluminum, the CNM evaporator processes additional metals.

Altogether, the different variety of designs and fabrication conditions account for potential systematic variability on $\lambda$, ensuring a more comprehensive analysis.

%
\section{Experiment}
\label{sec:exp}
\begin{figure}[!hbt]
    \centering
    \includegraphics{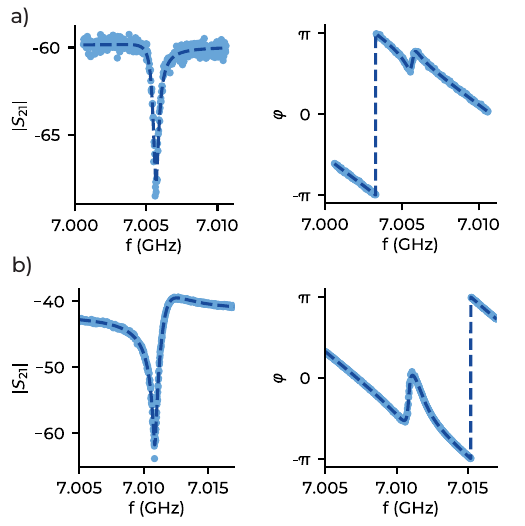}
    \caption{\(S_{21}\) resonator response in magnitude (\(|S_{21}|\), left) and phase (\(\varphi\), right). The fit using our procedure is superimposed to the light blue data points as a dark blue dashed line. \textbf{a)} \(100~\unit{\nm}\)-thick MR resonator. \textbf{b)} \(200~\unit{\nm}\)-thick MR resonator, with the same design as the device in \textbf{a)}.
    }
    \label{fig:fit}
\end{figure}

\begin{table*}[t]
    \caption{Results of the fits from Fig.~\ref{fig:fit}. All parameters are used in Eq. \ref{eq:s21} to fit the resonator response. Parameter uncertainties are obtained using iminuit, as detailed in the main text. $|Q_c|$ and $Q_l$ are related through $Q_l^{-1}=Q_i^{-1}+Re(Q_c^{-1})$ \cite{probst}.}
    \label{tab:fit-res}
    \begin{ruledtabular}
        \begin{tabular}{ccccccccccccccc}
            Res. ID & $f_r$ (\unit{\GHz}) & $\delta f_r$ (\unit{\kHz}) & $|Q_c|$             & $\delta |Q_c|$   & $Q_l$              & $\delta Q_l$ & $\phi$  & $\delta \phi$ & $\alpha$ & $\delta \alpha$ & $a$                  & $\delta a$        & $\tau (\unit{\Hz})$  & $\delta \tau (\unit{\Hz})$ \\
            \hline
            $0$     & $7.0056$            & $8$                        & $1.52 \cdot 10^{4}$ & $3 \cdot 10^{2}$ & $9.2 \cdot 10^{3}$ & $200$        & $0.22$  & $0.02$        & $-3.78$  & $0.04$          & $1.01 \cdot 10^{-3}$ & $3 \cdot 10^{-6}$ & $6.97 \cdot 10^{-8}$ & $7 \cdot 10^{-11}$         \\
            $1$     & $7.0115$            & $1$                        & $3.47 \cdot 10^{3}$ & $2$              & $4.2 \cdot 10^{3}$ & $3$          & $-0.76$ & $0.001$       & $-2.87$  & $0.001$         & $8.17 \cdot 10^{-3}$ & $1 \cdot 10^{-6}$ & $6.99 \cdot 10^{-8}$ & $5 \cdot 10^{-14}$
        \end{tabular}
    \end{ruledtabular}
\end{table*}

All resonator devices were mounted in a sample holder at the mixing chamber plate of a dilution refrigerator ($T=20~\unit{\milli\K}$) and subjected to measurements using a Vector Network Analyzer (VNA) at estimated power levels near the single photon regime. The full measurement setup schematics can be found in \cref{app:exp-set}. 4-probe structures were either located in the same chip as the resonators or in a separate chip, in which case they were located either at the mixing chamber or at the still plate of the dilution refrigerator.

Resistance measurements were performed at room temperature and at \(4~\unit{\K}\) [see \cref{fig:fab-exp}d)]. Then, a continuous acquisition was run at a low current while sweeping the temperature around \(T_{C,\mathrm{bulk}}\sim1.2~\unit{\K}\) to reach the critical temperature of the sample. \(T_C\) was determined at the midpoint of the superconductor-normal transition where the resistance dropped by 50\%. More details on the extraction process of $T_C$ is given in Appendix C, including a summary plot of all estimated values of $T_C$ obtained as function of thickness, which are also reported in Table II below.

A typical resonator measurement is shown in  \cref{fig:fab-exp}c) for two different resonator thicknesses. The resonator response was fit to the following expression \cite{probst}
\begin{equation}
    \label{eq:s21}
    S_{21} = a e^{i\alpha} e^{-2\pi i f \tau} \left[ 1 - \frac{(Q_l/|Q_c|)e^{i\phi}}{1+2iQ_l(f/f_r - 1)} \right] \, ,
\end{equation}
where the factor between brackets is the actual resonator response, while the pre-factor in front accounts for the response of the rest of the circuit. $a$ is an attenuation constant, $\alpha$ represents a phase shift, $\tau$ is the electrical delay of the measurement line, and $f_r$ is the resonator frequency. $Q_l$ is the loaded quality factor related to the complex coupling quality factor $Q_c\equiv|Q_c|e^{i\phi}$ and internal quality factor $Q_i$ through $Q_l^{-1}=Q_i^{-1}+Re(Q_c^{-1}) $\cite{probst}. $\phi$ expresses the impedance mismatch. Typical fits to the resonator response are shown in \cref{fig:fit}a)-b), where one can see that the impedance mismatch is correctly captured.

Our fitting process follows from a routine inspired by ref.~\cite{probst}. In that study, the data fitting is performed sequentially, until the full expression for \(S_{21}\) is obtained. However, with such a sequential process it is hard to propagate errors in the fitting, and correlations between parameters are lost. For that reason, we use the sequential procedure to yield our initial parameter guess. 
Once an initial set of parameters is obtained, we use the iminuit library \cite{iminuit} for fitting the full $S_{21}$ expression. Iminuit is a library mainly used by the particle physics community, with a focus on error propagation and accurate uncertainty estimation. The \(S_{21}\) fit has seven parameters, which makes it a complex fit, and proper error estimation becomes relevant. In most fits, the initial guess and final fitting parameters were close. In general, having good initial parameters for iminuit is indispensable. In some resonators, however, iminuit fitting found a considerably better solution than the initial one, while providing a correct error estimation.

\Cref{tab:fit-res} shows fitted parameters from resonators in \cref{fig:fit}, with the corresponding errors. The relative error in frequency $\delta f_r$ is of order 10$^{-6}$, which is the most important parameter to extract \(\lambda\). The resonator internal quality factor \(Q_i\) is not used in this study. In fact, the resonators in this work are designed with $Q_l\ll Q_i$ to maximize the signal strength, and therefore \(Q_i\) is not very accurately determined, lying in the range $10^3-10^4$ showing no obvious thickness dependence. Thickness dependence of $Q_i$ will be examined in future works.

%
\section{Results}
\label{sec:res}

\begin{figure}[hbt!]
    \centering
    \includegraphics{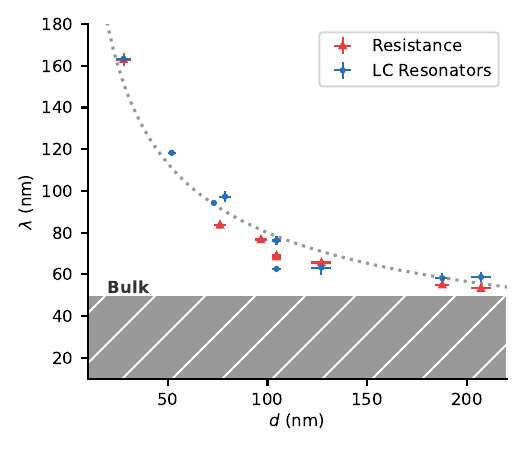}
    \caption{Measured penetration depth $\lambda$ as function of device thickness $d$. Two measurement methods are shown: resistance measurements (red triangles) and $LC$ resonator resonances (blue circles). Bulk penetration depth is shown as a lower limit. The thickness of samples is measured using atomic force microscopy. The fitting curve is $\lambda = a\lambda_L(\xi_0/d)^{1/2}$ using Eqs.~(\ref{eq:coh-len}, \ref{eq:pen-dep-dirty}) and assuming the electron mean-free path to be $l=d$. The fitted parameter is $a = 1.28\pm0.03$, half way between specular and diffusive scattering.}
    \label{fig:main-res}
\end{figure}

\begin{table*}[t]
    \caption{
    Results for both resistance and resonance frequency measurements. $d_{\mathrm{nom}}$ is the nominal target thickness, while $d_{\mathrm{meas}}$ is the measured thickness using atomic force microscopy. $\lambda$ is the measured average penetration depth. Measurement type is "LC" for lumped resonators and "R" for 4-probe resistance measurements. ``N" represents the number of devices measured for each particular thickness, which include small resonators (SR), medium resonators (MR), and large resonators (LR) for the resonance frequency measurements, and four-probe (4P) and Hall bar (HB) for the resistance measurements. The fabrication facility is specified for each device. For resonator measurements, $\lambda_{\mathrm{thin}}$ is shown, together with surface kinetic inductance $L_{k,s}$. For resistance measurements, residual-resistance ratio, defined as $RRR=R_{\mathrm{RT}}/R_{\mathrm{4K}}$ is shown.}
    \begin{ruledtabular}
        \begin{tabular}{cccccccccccccc}
            Device id & $d_{\mathrm{nom}} (\unit{\nm})$ & $d_{\mathrm{meas}} (\unit{\mathrm{\nm}})$ & $\lambda (\unit{nm})$ & Meas. & Design & N & Fab. & $\lambda_{\mathrm{thin}} (\unit{\nm})$ & $L_{k,s}$ ($\unit{\fH})$ & $RRR$        & $T_C (\unit{\kelvin})$ \\
            \hline
            1         & $25$                            & $28\pm 4$                                 & $163\pm1$         & LC    & MR     & 9 & IFAE & $965\pm12$                              & $1212\pm5$               & -            & -                      \\
            2         & $25$                            & $28\pm 4$                                 & $163\pm1$         & R     & 4P     & 1 & IFAE & -                                      & -                        & $2.1\pm0.1$  & $1.32\pm0.03$          \\
            3         & $50$                            & $52\pm 2$                                 & $118\pm1$         & LC    & MR     & 9 & IFAE & $286\pm4$                              & $359\pm2$                & -            & -                      \\
            4         & $75$                            & $73\pm 2$                                 & $94\pm1$          & LC    & MR     & 5 & IFAE & $145\pm3$                              & $182\pm3$                & -            & -                      \\
            5         & $75$                            & $76\pm 3$                                 & $84\pm1$          & R     & 4P     & 1 & CNM  & -                                      & -                        & $4.8\pm0.1$  & $1.22\pm0.01$          \\
            6         & $75$                            & $78\pm 3$                                 & $97\pm3$          & LC    & LR     & 9 & IFAE & $145\pm7$                              & $182\pm3$                & -            & -                      \\
            7         & $100$                           & $97\pm 3$                                 & $81\pm1$          & R     & 4P     & 1 & CNM  & -                                      & -                        & $6.1\pm0.1$  & $1.27\pm0.02$          \\
            8         & $100$                           & $105\pm 2$                                & $76\pm2$          & LC    & MR     & 5 & IFAE & $87\pm3$                               & $109\pm3$                & -            & -                      \\
            9         & $100$                           & $105\pm 2$                                & $63\pm1$          & LC    & LR     & 5 & IFAE & $67\pm1$                               & $84\pm3$                 & -            & -                      \\
            10        & $100$                           & $105\pm 2$                                & $69\pm1$          & R     & 4P     & 1 & IFAE & -                                      & -                        & $7.2\pm0.1$  & -                      \\
            11        & $100$                           & $105\pm 2$                                & $69\pm1$          & R     & HB     & 1 & IFAE & -                                      & -                        & $7.2\pm0.1$  & -                      \\
            12        & $100$                           & $127\pm 5$                                & $63\pm3$          & LC    & SR     & 9 & CNM  & $65\pm4$                               & $83\pm2$                 & -            & -                      \\
            13        & $100$                           & $127\pm 5$                                & $66\pm1$          & R     & 4P     & 1 & CNM  & -                                      & -                        & $7.4\pm0.1$  & $1.22\pm0.05$          \\
            14        & $200$                           & $188\pm 3$                                & $58\pm2$          & LC    & MR     & 5 & IFAE & $58\pm3$                               & $73\pm2$                 & -            & -                      \\
            15        & $200$                           & $188\pm 3$                                & $55\pm1$          & R     & 4P     & 1 & IFAE & -                                      & -                        & $11.6\pm0.1$ & $1.20\pm0.02$          \\
            16        & $200$                           & $207\pm 5$                                & $59\pm2$          & LC    & MR     & 6 & CNM  & $59\pm2$                               & $74\pm2$                 & -            & -                      \\
            17        & $200$                           & $207\pm 5$                                & $54\pm1$          & R     & 4P     & 1 & CNM  & -                                      & -                        & $9.9\pm0.1$  & $1.20\pm0.05$          \\
        \end{tabular}
        \label{tab:res}
    \end{ruledtabular}

\end{table*}

The obtained values of $\lambda$ for each sample thickness $d$ from both resonator and resistance measurements are plotted in \cref{fig:main-res}. Each point of the resonator measurement is obtained by an average of several resonators on the same chip with varying capacitance and kinetic inductance ratio (see \cref{tab:res}), as explained in Sec.~\ref{sec:sim}. The vertical error bars correspond to the standard error of the mean. The error in the horizontal axis corresponds to the uncertainty of the film thickness due to its granular nature. The resistance measurements are taken on a single structure in each device, with the vertical error bar corresponding to the error in the fitting of the resistance [Fig.~\ref{fig:fab-exp}]. The complete error propagation analysis is detailed in Apprendix~\ref{app:error}.

\Cref{tab:res} shows the full set of experimental results obtained for both the resonance and resistance measurements. A total of \(70\) structures have been analyzed, combining the two measurement methods (see Sec.~\ref{sec:mod}), three different designs for each measurement method, along with eight different thicknesses and two evaporation facilities. This combination reduces considerably the possibility of a systematic error on the experiment, and provides a good statistical ensemble from which to extract conclusive values of $\lambda$.

The data in \cref{fig:main-res} exhibit a consistent trend, with $\lambda$ increasing rapidly for $d<100~\rm{nm}$ and approaching bulk values for $d\sim200~\rm{nm}$, validating the chosen thickness range. For very low thicknesses, it is common to assume that the electron mean-free path $l$ is limited by surface scattering, leading to $l=d$~\cite{tinkham}. With this assumption, $\lambda$ can be fitted to Eq.~(\ref{eq:pen-dep-dirty}) leaving $a$ as a fitting parameter. The fit yields $a = 1.26\pm0.03$, which lies in a regime of mixed scattering. It is important to note that Eq.~(\ref{eq:pen-dep-dirty}) is only valid in the regime of a type-II superconductor, as the limit for large $l$ is $\lambda_L$, and not $\lambda_{\rm{bulk}}$. Using the fitted curve, $\lambda = \lambda_{\rm{bulk}}$ at $d\simeq250~\rm{nm}$, which sets this thickness as an upper limit to the validity of the expression.

\begin{figure*}[hbt!]
    \includegraphics{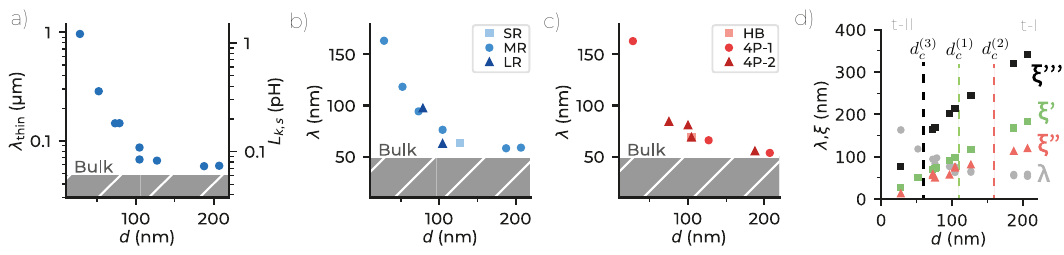}
    \centering
    \caption{\textbf{a)} $LC$ resonator response, plotted in two scales: \(\lambda_{\mathrm{thin}}\) (left y-axis) and $L_k$ (right y-axis), which are proportional, see Eq.~(\ref{eq:kin-ind-thin}). \textbf{b)} Measured $\lambda$ by the resonance frequency method for different thicknesses. Different resonator types are shown. The devices fabricated in CNM evaporator are highlighted. \textbf{c)} Measured $\lambda$ using the resistance method for different thicknesses. Different resistance structures are shown, the two types of four-probe, with (1) and without (2) ground plane, and the Hall bar. The devices fabricated in CNM evaporator are highlighted. \textbf{d)} Estimated dependence of coherence length $\xi$ on thickness using the fit from Fig.~\ref{fig:main-res}. $\xi'$ is calculated assuming $l=d$ with Eq~(\ref{eq:coh-len}). According to this fit, the transition between superconductivity regimes, when $\xi/\lambda=\sqrt{2}$~\cite{abrikosov}, occurs at a critical thickness $d_c^{(1)}\simeq113~\unit{nm}$. $\xi"$ is calculated from $l$ obtained from the relation $\rho \cdot l = 4\cdot 10^{-12}~\rm{\Omega \cdot  cm^2}$, leading to $d_c^{(1)}\simeq155~\unit{nm}$. Finally, $\xi'''$ used the $l(d)$ relation obtained from \cite{reale}, leading to $d_c^{(3)}\simeq55~\unit{nm}$.}  
    \label{fig:more-res}
\end{figure*}

The resonance frequency method gives access to $L_k$ through \cref{eq:lk-exp}, which is then converted to $L_{k,s}$ via \cref{eq:kin-sq-pen}. \(L_{k,s}\) is proportional to \(\lambda_{\mathrm{thin}}\), and both are shown in \cref{fig:more-res}a). In this figure, only $LC$ resonators are plotted, and the exponential increase at low thicknesses is more pronounced than the increase of $\lambda$ in \cref{fig:main-res}, as expected. Large $L_{k,s}$ values near \(\sim10~\unit{\pH}\) are predicted for aluminum with $d\sim10~\rm{nm}$. Combined with a long enough meander, pure aluminum can attain $L_k$ values comparable to those exhibited by superinductors~\cite{maleeva2018}.

Figures~\ref{fig:more-res}b) and c) show variations on the design and deposition system for both resonance and resistance measurements, respectively. The noticeable difference at $d=100~\rm{nm}$ may arise from a systematic difference in the device geometry between M and L resonators not accounted for.

Resistance measurements are displayed in \cref{fig:more-res}c). All values follow a consistent trend $\lambda\sim1/\sqrt{d}$. Most data are obtained from 4-probe structures with (4P-1) and without (4P-2) surrounding ground plane. The single Hall bar value is consistent with the rest of 4-probe measurements. Devices fabricated at different fabrication facilities did not show visible deviations in $\lambda$ despite exhibiting different residual resistance ratio $RRR$ values (\Cref{tab:res}).

The results obtained in this work allow us to provide estimates of the critical thickness $d_c$ where aluminum turns from a type-I to a type-II superconductor. Such a transition is expected to occur for a value of $\kappa = 1/\sqrt{2}$ in Eq.~(\ref{eq:kappa})~\cite{abrikosov}. In \cref{fig:more-res}d), the values of $\lambda$ obtained are compared to values of $\xi$ obtained from two different assumptions on the dependence of the electron mean-free path $l$ as function of thickness $d$. Our first method assumes $l=d$, which is a reasonable approximation for thin films as already argued, leading to a value $\xi'$ estimated from Eq.~(\ref{eq:coh-len}). The thickness that satisfies $\xi = \sqrt{2}\lambda$ is $d_c^{(1)}\simeq113~\rm{nm}$. It is important to note that this fit, which assumes a linear dependence between \(d\) and \(l\), fails for large $d$. An alternative method to predict $\xi$ is obtained by using the relation\cite{Brandt1971, Romijn} $\rho \cdot l = 4\cdot 10^{-12}~\rm{\Omega \cdot  cm^2}$ known for aluminum. According to this relation, a different $\xi"$ is obtained through Eq.~(\ref{eq:coh-len}) and the $\rho$ values obtained from the meanders, leading to $d_c^{(2)}\simeq 155~\rm{nm}$. In \cite{reale}, an additional relationship is proposed between $l$ and $d$ as
\begin{equation}
    \label{eq:l-d-reale}
    l = \frac{3}{4}\, \frac{1+p}{1-p} \, d \, \left( \log\frac{l_l}{d} + 0.4228 \right)\, ,
\end{equation}
where $p=0.08$ is the ratio of the charge carriers incident on the film surface which are scattered elastically, and $l_l=18.7~\unit{\um}$ is a corrected bulk mean free path \cite{gurzhi}. Using this expression, we obtain a third predicted critical thickness, $d_c^{(3)}\simeq 55~\rm{nm}$. The lower value of $d_c$ may be the result of inaccurate values of $p$ and $l_l$ which may differ for our films. In either case, an exact description of $l$ as function of $d$ is outside of the scope of this work.

Therefore our assumptions allow us to define a region of critical thicknesses $d_c = 55~\rm{nm}-155~\rm{nm}$ where the transition between superconductivity types may occur. As argued in Sec.~\ref{sec:theory}, the actual transition between superconductivity types may take place over a certain range of values of $d_c$, given the intrinsic inhomogeneity of $\kappa$ due to the disordered nature of thin films. In summary, our results suggest that the thicker samples ($d>150~\rm{nm}$) may enter the regime towards type-I superconductivity~\cite{Brandt1971}, unlike the thinner ones ($d<50~\rm{nm}$) which are more likely to behave as type-II local superconductors.

%
\section{Conclusions}
\label{sec:conclussions}

In this work, we performed a characterization of the aluminum penetration depth $\lambda$ for thicknesses in the range 28~nm--207~nm. The values of $\lambda$ obtained range from 163~nm for the thinnest samples down to 54~nm, approaching the aluminum bulk value for the thickest films.
Our measurements of $\lambda$ as function of thickness represent a guide to superconducting circuit designs where inductance plays a significant role, such as, e.g., circuit QED and kinetic inductance detectors.

We have also introduced an accurate fitting procedure for resonators which can easily be extended to other experimental settings. For simultaneous parameter fits, sequential fitting can be a good process to obtain each parameter at a time. However, this process does not provide a good estimation of the fitting error and the correlations between parameters. Using the result of sequential fitting as the initial guess, we then fitted with the iminuit package, providing an accurate error estimation and, in particular instances, a better parameter prediction.

Our work has also enabled us to provide an estimation of the thickness range at which thin-film aluminum starts behaving as a type-I superconductor. Further work is needed to explore the actual transition point and investigate potential new circuit functionalities in thicker aluminum films such as quasiparticle traps using the dependence of the superconducting gap on thickness \cite{pan, li2022} or the exclusion of vortices in thick enough films.

\section*{Data Availability statements}
The data that support the findings of this study are openly available at the following URL/DOI: [missing zenodo link]
%
\section*{Acknowledgements}
\label{sec:ack}
We would like to give special thanks to Sonnet support software for their help in running our simulations. We also want to thank Qilimanjaro engineers David Eslava and Yifei Chen for their collaboration and help during measurements, and Prof. Teun Klapwijk and Prof. Teresa Puig for their insightful comments. D.~L.-N. acknowledges support by an FPI grant from the Spanish Ministry of Science and Innovation (MICIN). We also acknowledge funding from the Ministry of Economy and Competitiveness and Agencia Estatal de Investigacion (RYC2019-028482-I, RyC-2026-21412, PCI2019-111838-2, PID2021-122140NB-C31, PID2021-122140NB-C32), the European Commission (FET-Open AVaQus GA 899561, QuantERA), and program ‘Doctorat Industrial’ of the Agency for Management of University and Research Grants (2020 DI 41; 2020 DI 42). The IMB-CNM is supported by the Mar\'ia de Maetzu grant for Centres of Excellence programme CEX2023-001397-M funded by MICIU/AEI/10.13039/501100011033. This work used the Spanish ICTS Network MICRONANOFAB. IFAE is partially funded by the CERCA program of the Generalitat de Catalunya. This study was supported by MICIN with funding from European Union NextGenerationEU~(PRTR-C17.I1) and by Generalitat de Catalunya.

\bibliographystyle{ieeetr}
\bibliography{citations.bib}

\newpage
\clearpage
\newpage
\appendix

%
\section{Experimental Setup}
\label{app:exp-set}
\begin{figure}[h!]
    \centering
    \includegraphics{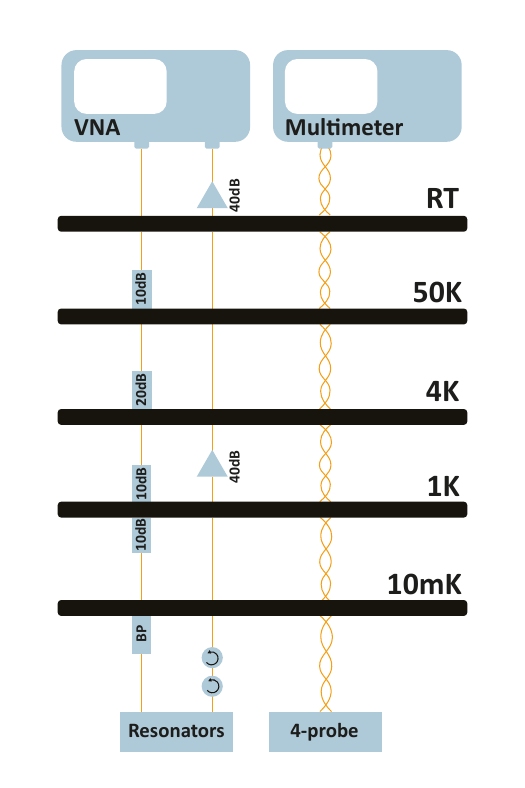}
    \caption{Measurement setup. High-frequency coaxial lines are attenuated 50dB and filtered above $8~\unit{GHz}$. The output signal is passed through two circulators and amplified through a HEMT amplifier and a room temperature amplifier. Resonators were measured with an Agilent Vector Network Analyzer E5071B. DC lines for resistance measurement were twisted pairs thermalized in several stages at the refrigerator. A Keithley 2600B source measurement unit was used for these measurements.}
    \label{fig:setup}
\end{figure}

The experimental setup used in the measurements of all devices presented in this work is shown in Fig.~\ref{fig:setup}.

%
\section{Fabrication}
\label{app:fab}

\begin{figure*}[t!]
    \centering
    \includegraphics{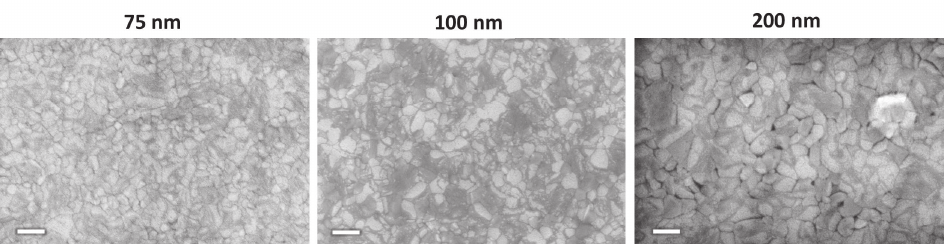}
    \caption{SEM Images for samples of increasing thickness illustrating the evolution in apparent grain size. The id of samples with 75~nm,
100~nm, and 200~nm correspond to 4, 9, 14, respectively. Scale bar corresponds to 200 nm.}
    \label{fig:grain}
\end{figure*}

High resistivity silicon wafers from TOPSIL provider are used as substrates for all the fabricated devices. The basic processing steps for patterning the electronic devices are as follows.

Samples are diced as quarter wafers and subsequently cleaned in acetone, rinsed in isopropanol and blown dry in N$_2$ stream. Substrates are dehydrated before photoresist deposition. LOR3A plus HIPR-6512 resists are spin coated and soft baked sequentially. A Karl Suss mask aligner is used for resist-stack exposure and then a development sequence using ODP is applied. The pattern transfer consists in aluminum thin film deposition followed by a lift off of the resist in NMP solvent. The aluminum deposition was performed by e-beam evaporation for the given thicknesses. Figure~\ref{fig:grain} shows SEM images of different samples to observe the grain size. The remaining LOR3A is removed by mild oxygen plasma treatment. For electrical characterization, the samples were diced appropriately and mounted onto ceramic PCBs. Devices were wire-bonded with aluminum wires to either a sample holder for AC measurements or to chip carriers for DC measurements. Exemplary images of actual devices are shown in Figs.~\ref{fig:fab-exp}a)-b).

\section{$T_C$ measurements}

\begin{figure}[t!]
    \centering
    \includegraphics{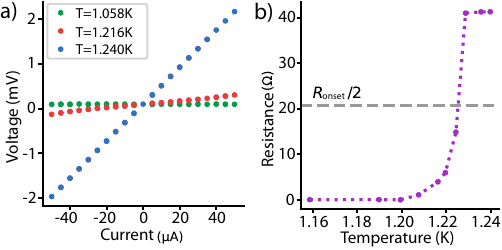}
    \caption{a) $I-V$ curve taken at different temperatures for a $127~\rm{nm}$-thick sample. b) Fitted resistance at different temperatures. The horizontal dashed line marks the value where resistance drops by half, $R_{\rm{max}}/2$, where $T_C$ is defined.}
    \label{fig:res-temp}
\end{figure}

In order to characterize the critical temperature $T_C$ of each sample measured in this work, $I-V$ curves are taken while the fridge temperature is swept in a controllable fashion. \cref{fig:res-temp}a) shows $I-V$s from three different temperatures for a sample $127~\rm{nm}$ thick. The temperature is recorded simultaneously at each data point of the $I-V$ curve. The value of the temperature used in the plots is then the average temperature across all the values of each $I-V$ curve. This way we take into account possible heating effects due to the largest currents applied. A simple linear regression is used to obtain the resistance $R$. The results of fitted resistance as function average temperatures is shown in \cref{fig:res-temp}b). The $R_{\rm{onset}}/2$ marks the value at which $T_C$ is defined, being a drop in resistance by a factor 2 from the temperature at which resistance starts to decrease suddenly.

\begin{figure}[t!]
    \centering
    \includegraphics{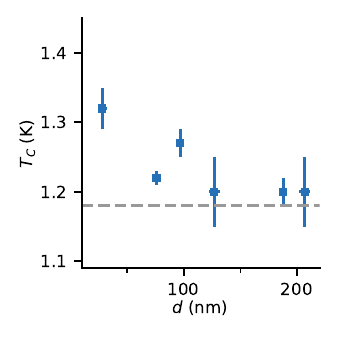}
    \caption{Critical temperature $T_C$ of resistor samples at different thicknesses, as measured following the procedure described in the text and in Fig.~\ref{fig:res-temp}.}
    \label{fig:tc}
\end{figure}

In \Cref{fig:tc}, the $T_C$ dependence with sample thickness is shown. As expected, $T_C$ is enhanced for lower thicknesses and approaches the bulk aluminium value for the thicker films, $T_C=1.18~\textrm{K}$.

%
\section{Simulations}
\label{app:sim}

\begin{figure}[!hbt]
    \centering
    \includegraphics{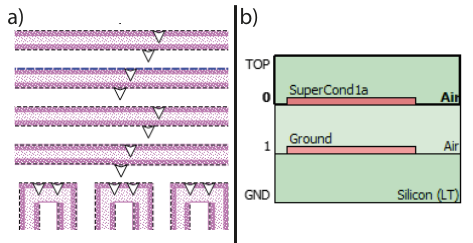}
    \caption{a) Sonnet design to simulate thickness in resonator. a) Zoom-in of the simulation design in a device like the one shown in Fig.~\ref{fig:sim}d), with the through vias used are represented as white triangles in the fingers (upper wires) and meanders (bottom turning wires) to connect the two layers of metal used to emulate thickness. b) Sonnet layer list, where two metallic layers are placed to simulate the film thickness.}
    \label{fig:ap-sim}
\end{figure}

FastHenry simulations were used to obtain an approximation of the kinetic inductance fractions of the resonator inductances, $\alpha\equiv L_k/L$. Those simulations determine the width and gap of the meanders, along with the number of meander turns. 

For FastHenry simulations, the value of \(\lambda\) had to be estimated, since it is needed for estimating the kinetic inductance $L_k$. We fitted a phenomenological behaviour of \(\lambda\)  as a decaying exponential with base $\lambda_{\mathrm{bulk}}=50~\unit{nm}$ and known values from previous internal studies at $d=25~\unit{nm}$ and $d=50~\unit{nm}$. The used expression is $\lambda(d) [\unit{\nm}] = \lambda_{\mathrm{bulk}} + 196 \cdot e^{-d[\unit{\nm}]/74.3}$.

With the final results, setting $4~\unit{\um}$ as the meander gap and width (which corresponds to MR-type resonators in \cref{sec:sim}), for $25~\unit{nm}$ thickness $\alpha=0.47$ and for $200~\unit{nm}$, $\alpha=0.03$, being the limits in thickness established for our work. LR- and SR-type resonators are designed for lower and higher thicknesses, respectively.

The main simulations were run with Sonnet, with $1~\unit{\um}$ fine meshing. To obtain different resonators with different resonance frequency but the same value of $L_k$, only the last fingers of the inter-digitated capacitor were modified [see main text \cref{fig:dev-des}b)], to keep the total inductance as constant as possible. The current in the last fingers of the capacitor is minimal [see main text \cref{fig:sim}d)]. This was easily achieved by parameterizing a Sonnet file and sweeping over the length of the last fingers. The frequency values obtained were approximately evenly spaced. The number of fingers modified and the distance between fingers was different for each resonator type (M, S and L).

When performing the simulation of a given resonator, not only the resonance frequency $f_0$ was acquired, but also the total inductance, $L$, and total capacitance, $C$. We followed the method described by Doyle~\cite{doyle}, which consists on adding an extra known sheet inductance in a single square of the meander and obtaining a new modified frequency $f_0'$. Together with $f_0$ and the known added inductance value, $L$ and $C$ can be separately obtained. More details can be found in the main text \cref{ssec:mod-res}.

Finally, we need to add the thickness into the simulation. This is a sensitive parameter in LC resonators, as shown in Fig.~\ref{fig:sim}c) of the main text, where the difference between sheet metal (gray resonance) and the thickest sample ($200~$nm) was around $2\%$, comparable to the difference added due to $L_k$ in real experiments for the thickest samples. In order to carry out the simulations, we followed a technique available within Sonnet that allows one to mimic thickness into an otherwise planar geometry, including the effect of sheet inductance~\footnote{This is to be distinguished from the `Thick metal' model available within Sonnet where no sheet inductance can be defined.}. In order to implement this technique, two layers of superconductor are added on top of the silicon, separated by a distance \(d\). Then, these two layers are connected through vias along all the edges. This effectively creates a thick resonator, reducing the inductance and increasing the capacitance. This technique is significantly faster to simulate entire resonator circuits as compared to using full 3D finite element solvers. A zoom-in of the simulation design we developed, including the vias connecting the vertical dimension, can be seen in \cref{fig:ap-sim}a-b). 

\begin{figure}[!hbt]
    \includegraphics{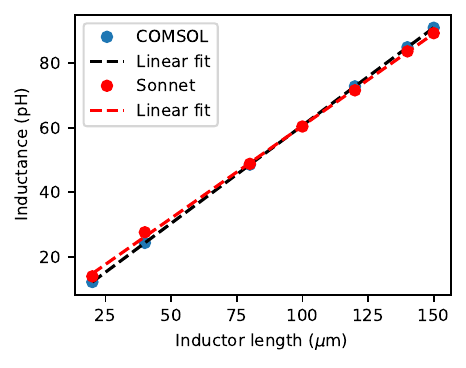}
    \caption{Simulated values of an inductor 4~$\mu$m wide, 100~nm thick using a full 3D model (COMSOL) and the thick-metal technique with Sonnet. Linear fits are added for visual aid. Both methods yield similar results with differences in the 1-2\% level.}
    \label{fig:ind-comp}
\end{figure}

We have validated the technique by simulating a linear inductor of the dimensions of our resonators using both Sonnet as well as the AC/DC-package from COMSOL. The inductor dimensions are 4~$\mu$m wide, 100~nm thick, and a varying length. We first run COMSOL simulations of the inductor to determine a suitable set of parameters. Using the 'Finer' meshing setting the error induced in the value of the inductance is below 0.15\%. We also find that a layer of 200~$\mu$m of Si substrate and 200~$\mu$m of air above the chip lead to sufficient values with error below 0.1\% compared to thicker geometries. The Sonnet simulations are carried out with the same geometry of the inductor, substrate and air box, using the technique to connect two layers to simulate finite thickness. The meshing is set to 100~nm, based on the limit set by the accuracy of the width (see Sec.~\ref{app:fab}). The resulting inductances with both methods as function of length is shown in Fig.~\ref{fig:ind-comp}. The values differ from each other less than 2\% in the worst cases, therefore justifying our simulation method. The simulation time of Sonnet for this inductor is a few seconds while it takes a few minutes for COMSOL, making it an advantageous technique.

Despite being more efficient than a full 3D simulation, these thick-metal Sonnet simulations required a significant amount of computer memory. In order to optimize simulation time, instead of simulating all the resonators with all the different finger lengths, for each resonator width, we only simulated the resonators with the shortest and longest last finger and compared the difference with their respective sheet resonators with no thickness, obtaining a ratio of frequencies. We found that this ratio of frequencies differed less than $<0.1\%$ among the different resonator line widths. Therefore, we applied this ratio to the remaining 2-dimensional resonators. This way, we obtained values of the simulated thick metal with just simulating the device with no thickness.

%
\section{Data Fitting}
\label{app:fit}

The main procedure to fit the data using the iminuit package is already detailed in Sec.~\ref{sec:exp}. Using iminuit one has access to the correlation matrix between variables, which is defined as
\begin{equation}
    \label{eq:corr}
    \rho_{ij} = \frac{\sigma_{ij}}{\sqrt{\sigma_{ii}\sigma{jj}}} \, .
\end{equation}
Here, $\sigma_{ii}$ is the variance of variable $i$, and $\sigma_{ij}$ is the covariance between variables $i$ and $j$, if $i\ne j$.

When using the fitting formula for the transmission measurements $S_{21}$ Eq.~(\ref{eq:s21}), both quality factors, $Q_L$ and $|Q_c|$, show high correlation and, indeed, $Q_L$ is bounded by $Q_c$. Other variables that are highly correlated are $f_r$ with $\phi_0$. $\alpha$ and $\tau$ are instead fully correlated, given that both appear in the exponential pre-factor of $S_{21}$. In both cases, a slight change in one of them, provokes adaptations in the other coupled variable to minimize the cost function, which indicates correlation. One could, in principle, parameterize the model in an alternative way to \cref{eq:s21} to use purely independent variables. However, this parameterization will probably require the use of variables with little physical meaning, thus losing the benefits and information obtained from fitting the variables with actual physical meaning.

In \cref{tab:fit-cor}, correlations between parameters are shown for resonator in \cref{fig:fit}a).

\begin{table}[!hbt]
    \caption{Table with correlation for fitting of resonator on Fig.~\ref{fig:fit}a). Parameters can be seen in Eq.~(\ref{eq:s21}).}
    \label{tab:fit-cor}
    \begin{ruledtabular}
        \begin{tabular}{cccccccc}
                     & $a$  & $\alpha$ & $\tau$ & $Q_l$ & $|Q_c|$ & $\phi$ & $f_r$ \\
            $a$      & 1    & 0        & 0      & -0.3  & -0.3    & -0.1   & 0.1   \\
            $\alpha$ & 0    & 1        & 1      & 0     & 0       & -0     & 0     \\
            $\tau$   & 0    & 1        & 1      & 0     & 0       & -0     & 0     \\
            $Q_l$    & -0.3 & 0        & 0      & 1     & 0.8     & 0      & -0    \\
            $|Q_c|$  & -0.3 & 0        & 0      & 0.8   & 1       & 0      & 0     \\
            $\phi$   & -0.1 & 0        & 0      & 0     & 0       & 1      & -0.7  \\
            $f_r$    & 0.1  & 0        & 0      & -0    & 0       & -0.7   & 1     \\
        \end{tabular}
    \end{ruledtabular}
\end{table}

%
\section{Error propagation}
\label{app:error}

In order to take into account the error introduced in the estimation of $\lambda$ due to the simulation of the resonator frequencies, we have performed a complete error propagation analysis.

The list of relevant equations already introduced in Secs.~\ref{sec:theory} and \ref{sec:mod} is the following:
\begin{equation}
\label{eq:lks}
L_{ks} = \mu_0\lambda_{\rm thin},
\end{equation}
\begin{equation}
\label{eq:l-thin}
\lambda_{\rm thin} = \mu_0 \lambda\coth(d/\lambda),
\end{equation}
\begin{equation}
\label{eq:lk}
L_k = NL_{ks},\,N=g/w,
\end{equation}
\begin{equation}
\label{eq:ff}
L_k = L_g\left(\frac{f_{\rm sim}^2}{f_{\rm meas}^2}-1\right).
\end{equation}
In order to obtain the error from $\lambda$ we first obtain the error from $\lambda_{\rm thin}$, as they are connected through an implicit relation, Eq.~(\ref{eq:l-thin}). Using Eqs.~(\ref{eq:lks}, \ref{eq:lk}, \ref{eq:ff}), $\lambda_{\rm thin}$ can be related to our measurable quantities: 
\begin{equation}
\label{eq:lthin}
\lambda_{\rm thin} = \frac{w}{\mu_0g}L_g\left(\frac{f_{\rm sim}^2}{f_{\rm meas}^2}-1\right),
\end{equation}
where the resonator inductor width $w$ resonator inductor length $g$ can be obtained by inspection of the devices using scanning electron microscopy, $L_g$ and $f_{\rm sim}$ can be obtained from finite-element simulation solvers like Sonnet, and, finally, $f_{\rm meas}$ is measured in the experiment with very high accuracy given by our sophisticated fitting method (see Appendix~\ref{app:fit}).

In estimating the errors in the observable quantities, we note that $\delta w\simeq\delta g$, but $\delta w/w\gg\delta g/g$. Therefore we can neglect the error coming from the length, $\delta g\simeq0$. Also, since our fitting method yields a very accurate value of the measured resonator frequency down to a few kHz, we consider its error negligible compared to the error in the simulated frequency, $\delta f_{\rm meas}\simeq0$. The error from $\lambda_{\rm thin}$ can be then computed by differentiation of Eq.~(\ref{eq:lthin}) with respect to $w$, $L_g$ and $f_{\rm sim}$:
\begin{multline}
\delta\lambda_{\rm thin} = \frac{1}{\mu_0g}\left(\frac{f_{\rm sim}^2}{f_{\rm meas}^2}-1\right)(L_g\delta w + w\delta L_g) +\\ \frac{2wL_g}{\mu_0g}\frac{f_{\rm sim}}{f_{\rm meas}^2}\delta f_{\rm sim}
\end{multline}
Re-arranging terms,
\begin{equation}
\lambda_{\rm thin}\left(\frac{\delta w}{w} + \frac{\delta L_g}{L_g}\right) + 2\left(\lambda_{\rm thin} + \frac{wL_g}{\mu_0g}\right) \frac{\delta f_{\rm sim}}{f_{\rm sim}}.
\end{equation}
Therefore, adding the errors in quadrature leads to
\begin{multline}\label{eq:dlthin}
\frac{\delta\lambda_{\rm thin}}{\lambda_{\rm thin}} = \left(\left[\left(\frac{\delta w}{w}\right)^2 + \left(\frac{\delta L_g}{L_g}\right)^2\right] +\right.\\ \left.4\left(\lambda_{\rm thin} + \frac{wL_g}{\mu_0g\lambda_{\rm thin}}\right)^2 \left(\frac{\delta f_{\rm sim}}{f_{\rm sim}}\right)^2\right)^{1/2}.
\end{multline}

The next step is to calculate the error from $\lambda$. To do that we use the relation between observed quantities ($\lambda_{\rm{thin}}$, $d$) and inferred ($\lambda$) as follows:
\begin{equation}
\lambda_{\rm{thin}} = \lambda\coth{\frac{d}{\lambda}}.
\end{equation}
The error on $\lambda_{\rm{thin}}$ consists of errors on $\lambda$ and $d$, which can be obtained by differentiating the expression with respect to each variable:
\begin{equation}
\delta\lambda_{\rm{thin}} = \left[\coth(d/\lambda) + \frac{d/\lambda}{\sinh^{2}(d/\lambda)}\right]\delta\lambda - \frac{\delta d}{\sinh^{2}(d/\lambda)} .
\end{equation}
Isolating $\delta\lambda$:
\begin{multline}
\delta\lambda = \frac{1}{\coth(d/\lambda)+(d/\lambda)\sinh^{-2}(d/\lambda)}\left[\delta\lambda_{\rm{thin}}\right.+\\\left.\frac{\delta d}{\sinh^{2}(d/\lambda)}\right].
\end{multline}
In adding them in quadrature, the total error in $\lambda$ is
\begin{multline}\label{eq:dlambda}
\delta\lambda = \frac{1}{\coth(d/\lambda)+(d/\lambda)\sinh^{-2}(d/\lambda)}\left[\delta\lambda_{\rm{thin}}^2+\right.\\ \left.\frac{\delta d^2}{\sinh^{4}(d/\lambda)}\right]^{1/2}.
\end{multline}
Equations~(\ref{eq:l-thin}, \ref{eq:lthin}, \ref{eq:dlthin}, \ref{eq:dlambda}) are used to calculate the values of $\lambda_{\rm thin}$, $\lambda$ and their respective errors in Table~\ref{tab:res} and Figs.~\ref{fig:main-res},~\ref{fig:more-res} of the main text.

In order to estimate the error from $\lambda$, we need to establish bounds on the errors from the width $\delta w$, thickness $\delta d$, resonance frequency $\delta f_{\rm sim}$, and the simulated geometric inductance $\delta L_g$.

\begin{figure*}
\includegraphics{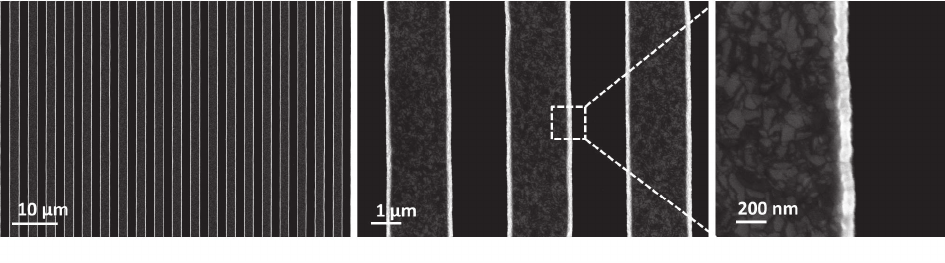}
\caption{Scanning Electron micrograph image of a section of the meander of resonator from device id 8 used in the experiment.}
\label{fig:res-w}
\end{figure*}

\begin{figure}[!hbt]
\includegraphics{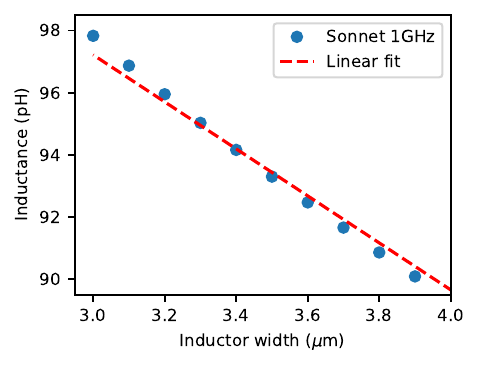}
\caption{Sonnet simulations of the width dependence of an inductor 150~\unit{\um} long, 100~nm thick. The slope around 4~\unit{\um} is -7.65pH/\unit{\um}.}
\label{fig:ind_ws}
\end{figure}

\begin{figure}[!hbt]
\includegraphics{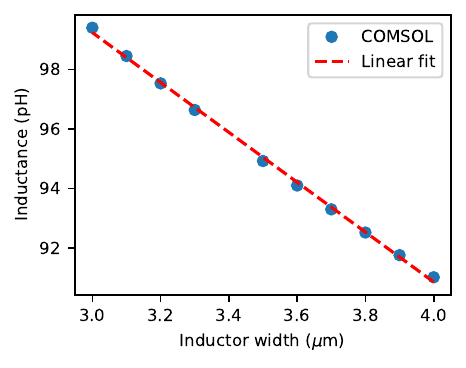}
\caption{COMSOL simulations of the width dependence of an inductor 150~\unit{\um} long, 100~nm thick. The slope around 4~\unit{\um} is -7.52pH/\unit{\um}.}
\label{fig:ind_wc}
\end{figure}

The bound on width fluctuations has been set to 200~nm based on scanning electron microscope images we have taken from several resonators. Figure~\ref{fig:res-w} shows an example from the meander of resonator from device id 8. Based on the variation of the width, we set $\delta w = 200$~nm as an upper bound. This variation is most likely limited by the resolution of the optical lithography and deposition processes used to fabricate the resonators. 
This limit on $\delta w$ is used to set a limit on the error from the simulated resonance frequency, $\delta f_{\rm sim}$. By using Sonnet on a simulated resonator circuit and modifying the width of the meander by 100~nm, the resonance is seen to vary on average by 0.6\%. Therefore we take $\delta f_{\rm sim}/f_{\rm sim} = 0.6\%$. Similarly, the uncertainty in the width is used to set the uncertainty in the value of the simulated geometric inductance, $\delta L_g$. In order to obtain the sensitivity of width-to-inductance, we simulated an inductor with length (150~\unit{\um}) and thickness (100~nm) comparable to that of our resonators of varying width around the nominal value of 4~$\mu$m, shown in  Figs.~\ref{fig:ind_ws},~\ref{fig:ind_wc} for both Sonnet and COMSOL. 
The dependence is slightly nonlinear, but the slope in both simulations differs by less than 2\% near 4~\unit{\um}, with a sensitivity of -7.65pH/\unit{\um} and -7.52pH/\unit{\um} for Sonnet and COMSOL, respectively. Longer meanders give a slightly higher sensitivity, up to 10pH/$\mu$m for 200~\unit{\um}-long meanders. Using this value as un upper bound we obtain a relative error in the inductor of $\delta L_g/L_g = 1\%$. The last error coming from the thickness is taken to be $\delta d=2~$nm, based on AFM images taken from several resonators. The error in the inductance from the uncertainty in the thickness is negligible, as the sensitivity of 16pH/\unit{\um} (found from simulations) is of the same order as the sensitivity of the width, but the magnitude of the uncertainty is much smaller.

\end{document}